\documentstyle[12pt,epsfig]{article}
\begin{document}
\title{
Optimization of the Parameters
in the RHIC Single Crystal Heavy Ion Collimation}
\author{
\underline{V.M.Biryukov}, Yu.A.Chesnokov, V.I.Kotov \\
{\small IHEP, Protvino, Russia}\\
D. Trbojevic,  A. Stevens \\ {\small BNL, Upton, USA} }
\date{Presented at the 1999 Particle Accelerator Conference, New York}
\maketitle
\begin{abstract}
In the framework of the project to design and test a collimation
system prototype using bent channeling crystal for
cleaning of the RHIC heavy ion beam halo, we have studied
the optimal length and bending angle of a silicon (110)
single crystal proposed to be a primary element situated
upstream of the traditional heavy amorphous collimator.
Besides the matters of the channeling and collimation efficiency,
we also looked into the impact the crystal may have
on the non-channeled particles that go on circulating in the
ring, so as to reduce the momentum offset of the particles
scattered of the crystal.
\end{abstract}

\section{INTRODUCTION}
The project of a collimation system using bent channeling
crystal for cleaning of the RHIC heavy ion beam halo has
been described in some detail in Ref.[1] where we refer
for the basic ideas and results, and more technical information.
In order to optimize the crystal collimation scheme
for RHIC, we have studied the influence of the silicon (110)
single crystal parameters (length, bending angle, curvature
gradient) on the efficiency of bending of the beam of fully
ionized Au ions. The beam distribution at the entrance
of the crystal was presented by the sample of 19258 fully
stripped gold ions generated as described in our previous
article [1].

We have also postulated that at the entrance to the crystal
all the particles have the same momenta, corresponding
to the Lorentz factor of 108.40, in order to evaluate then
the momentum distribution in the beam downstream of the
crystal.

In our first study [1] the crystal length of 10 mm and the
bending angle of 1 mrad were used, as it was a practical
choice corresponding to the state of art on that date. It was
concluded that the bending angle can be reduced substantially.
In the mean time, the IHEP Protvino experimental
practice has shown that shorter crystals of high quality can
be produced and they behave in perfect agreement with theory
[2].

\section{TRACKING IN CRYSTAL}
With the above said arrangements, we have performed
tracking of the 19258 particles through various crystals by
means of CATCH code [3]. Typical distribution of the particles
downstream of the crystal consists of a sharp peak
of the particles channeled through the full crystal length
and bent the full angle, some particles unbent but scattered
of the crystal, few particles bent partial angle between the
bent and unbent peaks, and few particles nuclear interacted
in the crystal. Fig. 1 represents gold ion distribution in the
horizontal phase space at the downstream edge of the crystal.

Having studied such distributions for the crystals with
different length but the same bending angle of 0.5 mrad,
we plot in Fig. 2 how many particles were bent at least 0.1
mrad (this includes the particles channeled part of the crystal
length as they are steered through the angles that might
be sufficient for interception by the downstream collimator).
One can see that, from standpoint of physics alone,
one can reduce the crystal length to nearly 2 mm without
seriously affecting the efficiency of bending of the above-defined
sample of particles.

Further we studied how the efficiency of the 5-mm long
crystal would depend on the bending angle in the range of
about 0.2-1.0 mrad. There was hardly any dependence up
to at least 1 mrad, so this gives us the freedom to adjust the
bending angle to what seems appropriate for collimation
purpose in RHIC.

One objective on the agenda was to find if we could benefit
in the efficiency from a variable curvature of the crystal
(in the above simulations the curvature was constant).
The motive is that, whereas in the extraction mode one has
to deliver the bent beam to a single direction, in the collimation
mode it is sufficient to deliver the bent beam to a
range of directions, say from 0.3 to 1.0 mrad in our case, as
the particles are absorbed anyway (it might be even useful
to spread the irradiation load over the collimator). In the
simulation we have tried variable curvature of the 5-mm
crystal, with the curvature peak being factor of 2, 4, 8, and
16 of the average. However, the fraction of the particles
bent at the angles greater than 0.1 mrad has not changed
significantly as compared to the above studies, so we don't
mention this option further on.

Table 1 summarizes comparative characteristics of the
beam interaction with two crystals, 5 mm long and 10 mm
long. The total number of the gold ions incident on the
crystals was $N$ particles. The number of particles which interacted
with the crystal nuclei was $N_{Lostparticles}$.The
$N_{channeled}$ represents the number of particles channeled
through the full length of the crystal and hence bent the full
angle, and Eff is their fraction to the total.

\begin{table}[htb]
\begin{center}
\caption{The results of beam interaction with two Si(110)
crystals.}
\begin{tabular}{|c|c|}
\hline
Crystal 1 cm& Crystal 0.5 cm \\
\hline
19258& 19258\\
962& 456\\
12532& 13686\\
17289& 18628\\
467& 174\\
65.1\%& 71.1\%\\
\hline
\end{tabular}
\end{center}
\end{table}

The momentum distribution has substantially improved
with shorter crystals as compared to the earlier study [1].
The $N_{inbucket}$ and $N_{outoffbucket}$ in Table 1 show the corresponding
number of the particles from the momentum
distribution downstream of the crystal. Figure. 3 shows
phase space of the gold ions at the end of the 5-mm long
crystal, where on the x coordinate is shown as dp/p dependent.
Taking into account also the practical considerations and
experience obtained with crystals at IHEP Protvino, we can
conclude that 5-mm long crystal seems well suited for the
job of Au beam bending at the angle of the order of 0.5
mrad in a single passage. Several crystals of this size have
been already produced and tested at Protvino [2].

One option for the future simulation studies may sound
interesting. Fig. 2 shows that from physics considerations
we can reduce the crystal length to order of 1-2 mm, if it
appears convenient technically. Such crystal would make
a rather little disturbance to the circulating beam, thus allowing
potentially many passages of the circulating particles
through the bent crystal and enhancing the overall efficiency
of bending. To benefit from this mode, such a crystal
should have also a sufficient transversal size (possibly
greater than the longitudinal size). Studies on feasibility of
such crystal deflectors are in progress.
\section{COLLIMATION SIMULATION IN RHIC}

The gold ions tracked through the crystal have been transported
through the RHIC ring using the tracking program
TEAPOT [4], with accelerator setting as described in Ref.
[1]. Fig. 4 shows the losses around the RHIC rings from
the particles scattered of the primary and secondary collimators
and the losses from the particles deflected by the
crystal. Two extreme cases are presented when the primary
collimator downstream of the crystal is wide open
and when it is set at 5 $\sigma_x$ , the same horizontal distance as a
front edge of the crystal.

\section{CONCLUSIONS}
An improvement in the momentum distribution of the ions
passed through the crystal and a reduction of lost particles
around the ring have been achieved by optimizing the crystal
length. Another result is the freedom we have in choosing
the angle of bending, which can be left about 0.5 mrad,
or be chosen from consideration of collimation physics, or
be just varied in the course of experimentation at RHIC.

\begin{figure}[h]
\begin{center}
\parbox[c]{13.5cm}{\epsfig{file=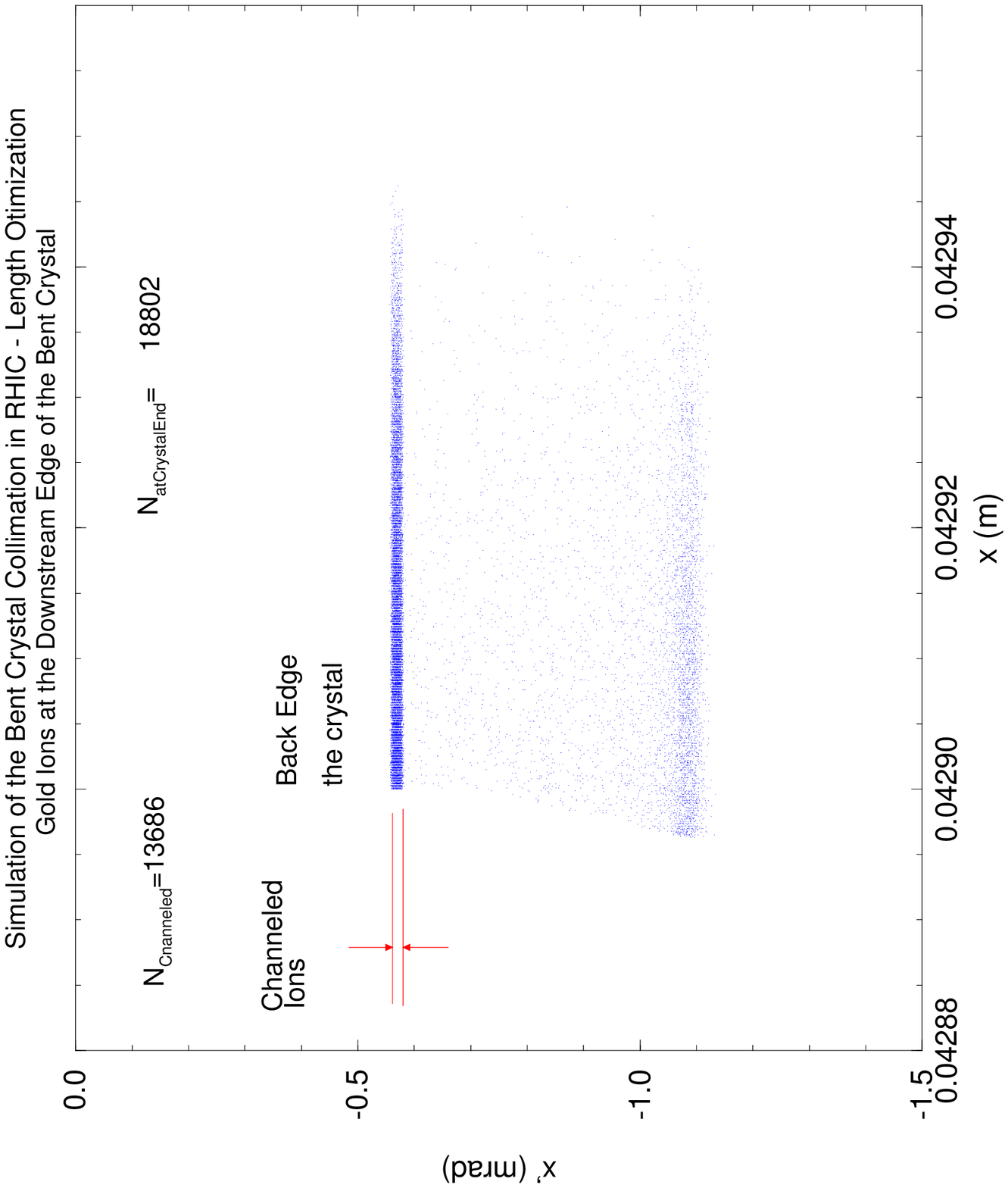,width=12.cm }}
\caption { Gold ion distribution in the horizontal phase
space at the downstream edge of the 5-mm long crystal.
 }
\end{center}
\end{figure}

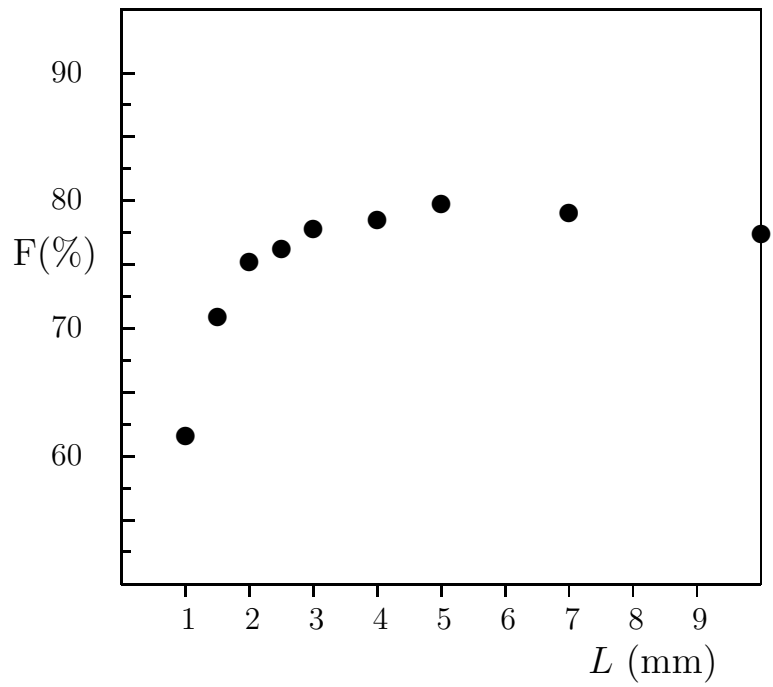
\begin{figure}[htb]
\begin{center}
\setlength{\unitlength}{.85mm}
\begin{picture}(110,100)(0,-10)
\thicklines
\linethickness{.25mm}
\put(    100.,54.8)  {\circle*{3}}
\put(    70.,58.)  {\circle*{3}}
\put(    50.,59.4)  {\circle*{3}}
\put(    40.,57)  {\circle*{3}}
\put(    30.,55.6)  {\circle*{3}}
\put(    25.,52.4)  {\circle*{3}}
\put(    20.,50.4)  {\circle*{3}}
\put(    15.,41.8)  {\circle*{3}}
\put(    10.,23.2)  {\circle*{3}}

\put(0,0) {\line(1,0){100}}
\put(0,0) {\line(0,1){90}}
\put(0,90) {\line(1,0){100}}
\put(100,0){\line(0,1){90}}
\multiput(10,0)(10,0){9}{\line(0,-1){2}}
\multiput(0,10)(0,10){8}{\line(1,0){2}}
\multiput(0,5)(0,5){16}{\line(1,0){1.4}}
\put(10,-7){\makebox(1,1)[b]{1}}
\put(20,-7){\makebox(1,1)[b]{2}}
\put(30,-7){\makebox(1,1)[b]{3}}
\put(40,-7){\makebox(1,1)[b]{4}}
\put(50,-7){\makebox(1,1)[b]{5}}
\put(60,-7){\makebox(1,1)[b]{6}}
\put(70,-7){\makebox(1,1)[b]{7}}
\put(80,-7){\makebox(1,1)[b]{8}}
\put(90,-7){\makebox(1,1)[b]{9}}
\put(-11,20){\makebox(1,.5)[l]{60}}
\put(-11,40){\makebox(1,.5)[l]{70}}
\put(-11,60){\makebox(1,.5)[l]{80}}
\put(-11,80){\makebox(1,.5)[l]{90}}

\put(-17,50){\large F(\%)}
\put(73,-14){\large $L$ (mm)}

\end{picture}
\end{center}
\caption{ Single-pass bending efficiency vs crystal length;
for 0.5 mrad bending angle.
}
  \label{a1}
\end{figure}

\begin{figure}[h]
\begin{center}
\parbox[c]{13.5cm}{\epsfig{file=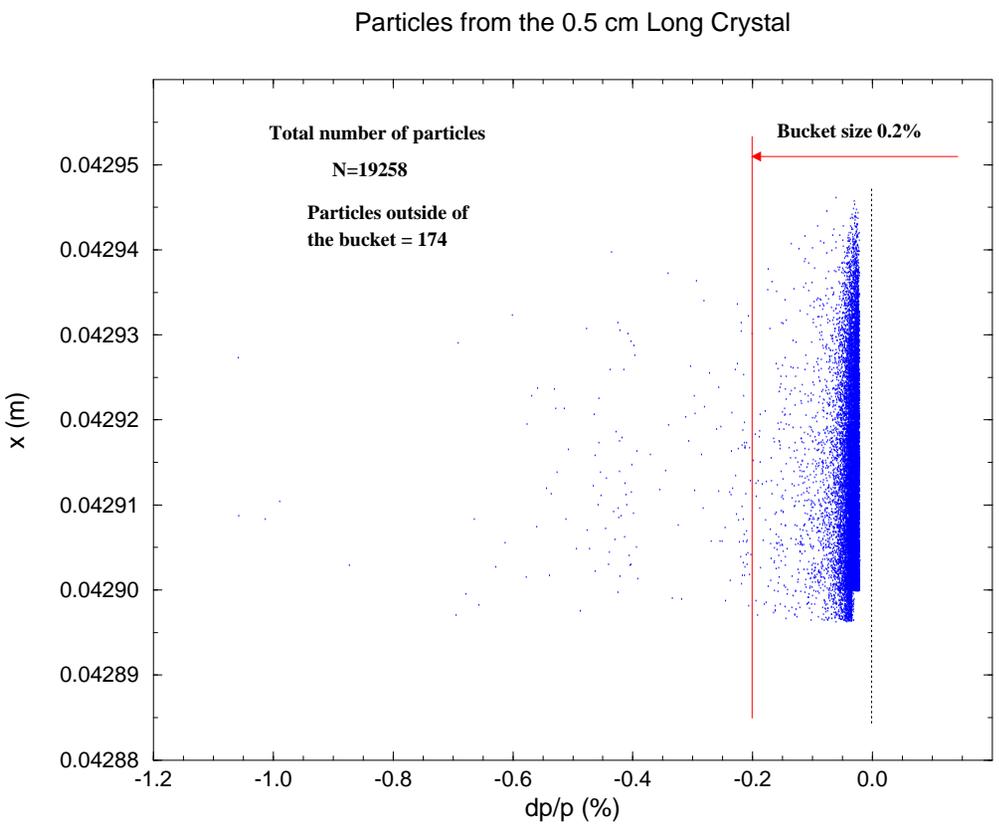,width=12.cm }}
\caption { Phase space of the gold ions at the end of the
5-mm long crystal.
}
\end{center}
\end{figure}

\begin{figure}[h]
\begin{center}
\parbox[c]{13.5cm}{\epsfig{file=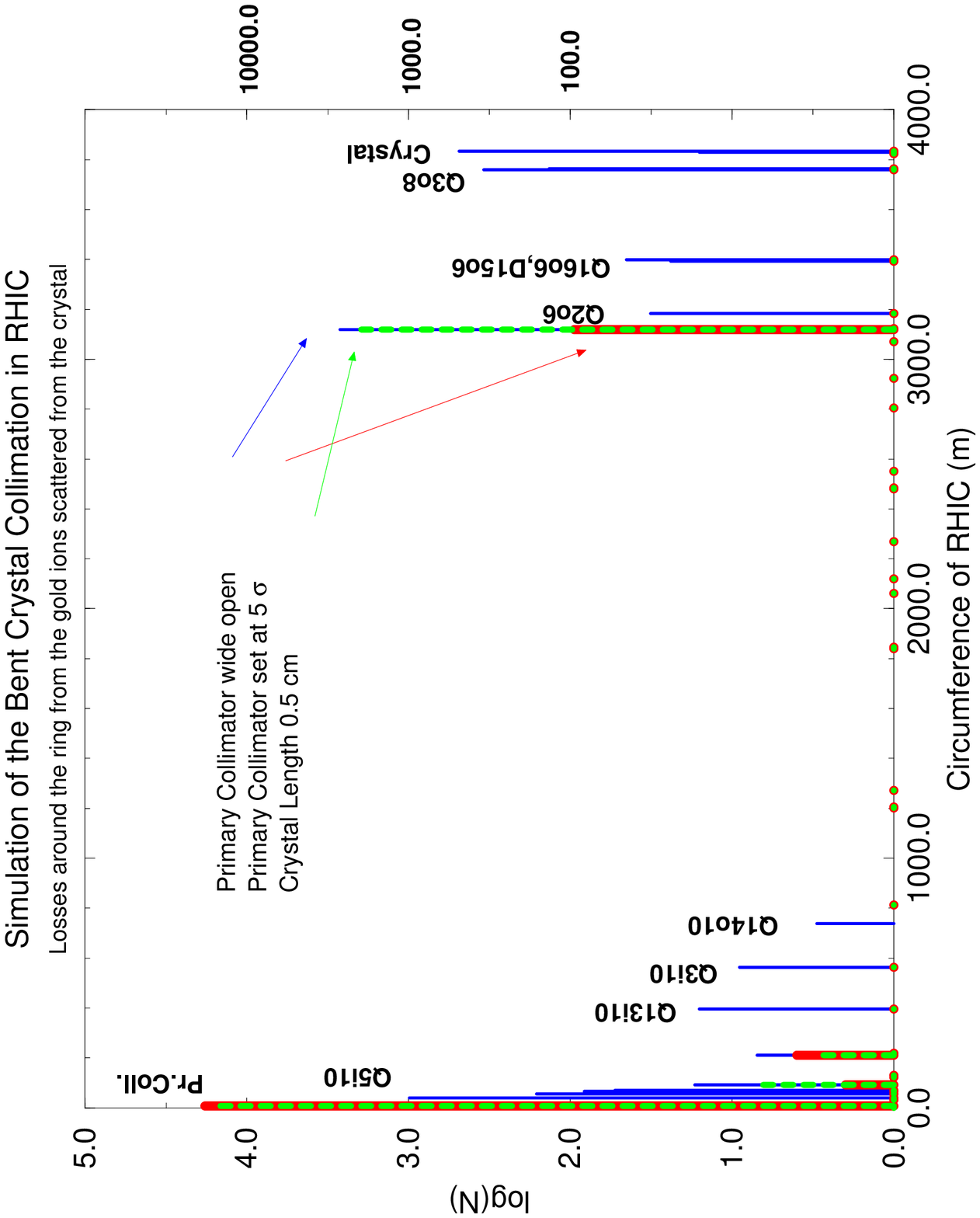,width=12.cm }}
\caption { Losses around the RHIC rings. }
\end{center}
\end{figure}


\begin{thebibliography}{**}

\bibitem{epac}
D.~Trbojevic {\it et al.},
"A Study Of RHIC Crystal Collimation,"
[arXiv:hep-ex/0111021]
in EPAC Proceedings (Stockholm, 1998) and references therein.
 V. Biryukov, Proceedings of the
``International Symposium on Near Beam Physics'', at Fermi National Laboratory, Batavia, Illinois, USA, Sptember 22-24, 1997, pp. 179-184 [arXiv:hep-ex/0110074].

\bibitem{plb}
A.~G.~Afonin {\it et al.},
JETP Lett.\  {\bf 68}, 568 (1998)
Phys.\ Lett.\ B {\bf 435}, 240 (1998).
JETP Lett.\  {\bf 67}, 781 (1998)
[arXiv:hep-ex/0111028].
\bibitem{3}
 Valery Biryukov, "Crystal Channeling Simulation - CATCH
1.4 User's Guide", SL/Note 93-74(AP), CERN, 1993.
V.~Biryukov,
Phys.\ Rev.\ E {\bf 51}, 3522 (1995).
Phys.\ Rev.\ E {\bf 52}, 2045 (1995).
Phys.\ Rev.\ E {\bf 52}, 6818 (1995).
Phys.\ Rev.\ Lett.\  {\bf 74}, 2471 (1995).
\bibitem{4}
 L.Schachinger and R.Talman, "A Thin Element Accelerator
Program for Optics and Tracking", SSC Central Design
Group, Internal Report SSC-52 (1985).
\end{thebibliography}
\end{document}